\documentclass[aps,prl,reprint,showpacs,superscriptaddress]{revtex4-1}

\usepackage[T1]{fontenc}
\usepackage[utf8]{inputenc}
\usepackage{graphicx,color,rotating,pifont}
\usepackage{amsmath,amssymb}
\usepackage{mathtools}
\usepackage{siunitx}
\usepackage{bm}
\usepackage{ae}
\usepackage{dcolumn}
\usepackage{txfonts}
\usepackage{tensor}
\usepackage{braket}
\usepackage{booktabs}
\usepackage{xspace}
\usepackage{soul}
\setstcolor{red}
\setul{}{.2ex} 

\usepackage{hyperref}

\DeclareMathSymbol{\NS}{\mathord}{AMSb}{"4E}

\DeclareSIUnit{\fm}{\femto\meter}

\newcommand{\abs}[1]{\lvert {#1} \rvert}

\newcommand{\comm}[2]{\ensuremath{[{#1},{#2}]}}

\newcommand{\op}[1]{\ensuremath{#1}}

\newcommand{\dd}{\ensuremath{\mathrm{d}}}

\renewcommand{\vec}[1]{\ensuremath{\bm{#1}}}

\newcommand{\hw}{\ensuremath{\hbar\omega}}

\newcommand{\eMax}{\ensuremath{e_{\text{max}}}}

\newcommand{\nuc}[2]{\ensuremath{{}^{#2}\mathrm{#1}}}

\definecolor{FGViolet}{rgb}{0.61,0.32,0.61}
\definecolor{FGDarkBlue}{rgb}{0,0,0.6}
\definecolor{FGBlue}{rgb}{0,0,0.8}
\definecolor{FGLightBlue}{rgb}{0.2, 0.6, 0.8}
\definecolor{FGGreen}{rgb}{0.2,0.7,0.2}
\definecolor{FGLightGreen}{rgb}{0.4,1,0.4}
\definecolor{FGYellow}{rgb}{1,0.95,0}
\definecolor{FGOrange}{rgb}{0.95,0.5,0.1}
\definecolor{FGRed}{rgb}{0.8,0,0}
\definecolor{FGWhite}{rgb}{1,1,1}
\definecolor{FGLightGray}{rgb}{0.8,0.8,0.8}
\definecolor{FGGray}{rgb}{0.5,0.5,0.5}
\definecolor{FGDarkGray}{rgb}{0.3,0.3,0.3}
\definecolor{FGBlack}{rgb}{0,0,0}

\begin{document}

\title{\emph{Ab Initio} Treatment of Collective Correlations and the
Neutrinoless Double Beta Decay of \texorpdfstring{$^{48}$Ca}{48Ca}}

\author{J. M. Yao}
\email{yaoj@frib.msu.edu}
\affiliation{Facility for Rare Isotope Beams, Michigan State University,
East Lansing, Michigan 48824-1321}

\author{B. Bally}
\email{bbally@email.unc.edu}
\author{J. Engel}
\email{engelj@physics.unc.edu}
\affiliation{Department of Physics and Astronomy, University of North Carolina,
Chapel Hill, North Carolina 27516-3255, USA}

\author{R. Wirth}
\email{wirth@frib.msu.edu}
\affiliation{Facility for Rare Isotope Beams, Michigan State University,
East Lansing, Michigan 48824-1321}

\author{T. R. Rodr\'{\i}guez}
\email{tomas.rodriguez@uam.es}
\affiliation{Departamento de Física Teórica y Centro de Investigación Avanzada
en Física Fundamental, Universidad Autónoma de Madrid, E-28049 Madrid, Spain}

\author{H. Hergert}
\email{hergert@frib.msu.edu}
\affiliation{Facility for Rare Isotope Beams, Michigan State University,
East Lansing, Michigan 48824-1321}
\affiliation{Department of Physics \& Astronomy, Michigan State University,
East Lansing, Michigan 48824-1321}

\date{\today}

\begin{abstract}

Working with Hamiltonians from chiral effective field theory, we develop a novel
framework for describing arbitrary deformed medium-mass nuclei by combining the
in-medium similarity renormalization group with the generator coordinate method.
The approach leverages the ability of the first method to capture dynamic
correlations and the second to include collective correlations without
violating symmetries.  We use our scheme to compute the matrix element that
governs the neutrinoless double beta decay of $^{48}$Ca to $^{48}$Ti, and find
it to have the value $0.61$, near or below the predictions of most
phenomenological methods.  The result opens the door to \emph{ab initio}
calculations of the matrix elements for the decay of heavier nuclei such as
$^{76}$Ge, $^{130}$Te, and $^{136}$Xe.

\end{abstract}


\maketitle

\clearpage

\paragraph{Introduction.}
The discovery that neutrinos oscillate
\cite{Super-Kamiokande:1998,SNO:2001,KamLAND:2003,Dayabay:2012} and thus have
mass has increased the significance of neutrinoless double beta
($0\nu\beta\beta$) decay \cite{Furry:1939}, a hypothetical rare process in which 
a parent nucleus decays into a daughter with two fewer
neutrons and two more protons, while emitting two electrons but no
(anti)neutrinos. The search for this lepton-number-violating process has become
a priority in nuclear and particle physics; its observation would have
fundamental implications for the nature of neutrinos, physics beyond the
Standard Model, and cosmology. 

$0\nu\beta\beta$ decay can result from the exchange of heavy particles in
lepton-number violating theories, but whatever the cause, a nonzero decay rate
implies a contribution from the exchange of a light Majorana neutrino, and we
focus on that contribution here. If it dominates, the inverse $0\nu\beta\beta$
half life is given by 
\begin{equation}
\label{half-life}
 [T^{0\nu}_{1/2}]^{-1} = g^4_A G_{0\nu} \left\lvert\dfrac{\langle
 m_{\beta\beta}\rangle}{m_e}\right\rvert^2 \left\lvert M^{0\nu}\right\rvert^2\,,
\end{equation}
where $m_e$ is the electron mass, $g_A$ the axial-vector coupling, and $G_{0\nu}
\sim10^{-14}{\rm yr}^{-1}$ is a phase-space factor 
\cite{Kotila:2012,Stoica:2013,Simkovic:2018}. The effective Majorana
neutrino mass $\langle m_{\beta\beta}\rangle=\bigl\lvert\sum_k
U^2_{ek}m_k\bigr\rvert$ contains physics beyond the Standard model through the
masses $m_k$ and the elements $U_{ek}$ of the Pontecorvo-Maki-Nakagawa-Sakata
flavor-mixing matrix. Certain combinations of these parameters
have been measured, but the individual masses $m_k$ and the combination
$m_{\beta\beta}$ are still unknown.  Equation \eqref{half-life} provides a way
to determine $\langle m_{\beta\beta}\rangle$ from a measured half life (which
must be significantly longer than that of any other process ever observed) if
the nuclear matrix element (NME) $M^{0\nu}=\braket{F|\op{O}^{0\nu}|I}$ of the
decay operator $\op{O}^{0\nu}$ between the ground states of the initial
($I$) and final ($F$) nuclei is known.  (See Ref.\ \cite{Simkovic:2008} for the
precise form of $\op{O}^{0\nu}$.) Since the NME cannot be measured, 
it must be computed from theory.

Although calculating an NME is straightforward in principle, the values
predicted by nuclear models differ by factors of up to three, causing an
uncertainty of an order of magnitude (or more) in the half-life for a
given value of $m_{\beta \beta}$ \cite{Engel:2017}.  It is difficult to reduce
this uncertainty because each model has its own phenomenology and uncontrolled
approximations.  To avoid model dependence, several groups have begun programs
to calculate the NMEs from first principles, taking advantage of the progress in
nuclear-structure theory in recent decades
\cite{Navratil:2009, Lee:2009, Barrett:2013, Launey:2016, Soma:2014, Hagen:2014, 
Hergert:2016jk, Hergert:2016,Stroberg:2019th, Carlson:2015lq, Lynn:2019dw, 
Shen:2019}.  However, applying modern \textit{ab initio} 
methods to $\beta\beta$
decay poses a significant challenge.  The $0\nu\beta\beta$ candidate nuclei are
generally heavier and more structurally complicated than those treated so far,
and the NME entails a much more involved calculation than do matrix elements of
the Hamiltonian or other simple operators.  Recently, \emph{ab initio} quantum
Monte Carlo methods have been used to calculate NMEs \cite{Pastore:2018}, but
only in very light nuclei that are of no interest to experimentalists.  

Among the \textit{ab initio} methods that can be applied to heavier nuclei, the
in-medium similarity renormalization group (IMSRG)
\cite{Hergert:2016jk,Hergert:2016} seems particularly promising because its time
and memory requirements scale polynomially with the underlying single-particle
basis size, and depend only indirectly on the particle number $A$.  The IMSRG uses a flow equation
to gradually transform the Hamiltonian so that a preselected ``reference
state'' becomes the ground state. Like the similar coupled-cluster approach
\cite{Hagen:2014}, the method has thus far been applied only to spherical
nuclei, which require only relatively simple reference states. Here, we
extend the IMSRG to deformed nuclei by combining it with the generator
coordinate method (GCM) \cite{Griffin:1957,Ring:1980}, which successfully
describes nuclei with complex shapes in nuclear density functional theory (DFT)
\cite{Bender:2008, Yao:2010, Rodriguez:2010m}.  This innovation removes the
heavy burden of capturing collectivity from the IMSRG by using it in conjunction
with deformed reference states and subsequently full GCM wave functions rather
than single Slater determinants.
Importantly, it does so without introducing new phenomenology; our \textit{ab
initio} method starts from interactions derived from chiral effective field
theory (EFT), the parameters of which are fixed in the lightest nuclei, and
systematically approaches an exact solution of the Schr\"odinger equation as
approximations are removed.  We have already tested our new many-body approach in a
small shell-model space with a phenomenological Hamiltonian \cite{Yao:2018wq}
and in large spaces with a chiral Hamiltonian in nuclei with $A=6$
\cite{Basili:2019}.  Here, we compute the NME for the
$0\nu\beta\beta$ decay of $^{48}$Ca to the deformed nucleus $^{48}$Ti. This
particular decay is a natural starting point because $^{48}$Ca is the lightest
candidate for an experiment, and a program to use it already exists
\cite{Iida:2016}.


\paragraph{Methods.}%

The starting point for any \textit{ab initio} calculation is a Hamiltonian with
coupling constants fit to reproduce data in few-nucleon systems.  We
take ours from chiral EFT, which systematically organizes interaction terms
by powers of a ratio of a typical nuclear
momentum scale and a larger hadronic scale.  The fit of parameters for
the two-nucleon interaction, carried out at next-to-next-to-next-to leading
order (N$^3$LO) with a momentum cutoff of 500 MeV/$c$, is from Entem and Machleidt (EM)
\citet{Entem:2003}.  We use the free-space (not
in-medium) similarity renormalization group (SRG) \cite{Bogner:2010} to
transform the interaction so that it has a ``resolution scale'' of either
$\lambda=1.8$ or $\SI{2.0}{\per\fm}$.  Following Refs.\
\cite{Hebeler:2011,Nogga:2004il}, we construct the three-nucleon interaction
directly, with a chiral cutoff of $\Lambda=\SI{2.0}{\per\fm}$.  We refer to 
the resulting Hamiltonians as EM$\lambda$/$\Lambda$, e.g., EM1.8/2.0 (
see Supplemental Material and Refs.~\cite{Hebeler:2011,Nogga:2004il} for further
details).  

To make our Hamiltonians easier to use, we normal order the three-body
interaction with respect to an $A=48$ Slater-determinant reference state and
omit the residual three-body terms, in what is called the normal-ordered
two-body (NO2B) approximation.  We also completely drop all three-body matrix
elements involving states with single-particle energies $e_{i}$ (in units of
$\hbar \omega$, the harmonic-oscillator spacing for our working basis) that sum
to $e_1+e_2+e_3 > 14$.  We let $H(0)$ stand for all Hamiltonians generated by
these procedures.

Once we have an appropriate $H(0)$, we combine the IMSRG with the GCM to compute
properties of $^{48}$Ca and $^{48}$Ti, in particular their ground-state wave
functions and the NME between them.  Roughly speaking, this task has two steps.
The first, as mentioned above, is the construction of a transformed Hamiltonian
for which our chosen reference states are good approximations to the ground states.
We obtain these states by using $H(0)$ in particle-number projected
Hartree-Fock-Bogoliubov (HFB) calculations, with variation after projection
\cite{Ring:1980}.  This allows us to explicitly include collective
deformation and pairing correlations, which, as we noted earlier, are difficult
to generate in the particle-hole-like expansion underlying the IMSRG as
practiced so far \cite{Hergert:2016,Hergert:2018th,Parzuchowski:2017ta}.  The
IMSRG flow equation~\cite{Morris:2015} then creates a set of unitary
transformations $\op{U}(s) \equiv e^{\op{\Omega}(s)} = \mathcal{S} \exp
\int_{0}^{s'} \! \eta (s')\, \mathop{ds'}$ (an $s$-ordered exponential), where
$s$ is the flow parameter and $\eta(s)$ is a ``generator'' that is chosen to
make the reference states increasingly close to ground states as $s$ increases.
We employ the Brillouin generator \cite{Hergert:2016}, which produces steepest
descent in the energy Tr$[\rho H(s)]$, where $\rho$ is a density operator and we
use the notation that for any ``bare'' operator $O(0)$, $O(s) \equiv \exp[\Omega
(s)] O(0) \exp[-\Omega (s)]$ is the transformed operator at flow-parameter $s$.
To make the flow equation tractable, we truncate all such operators at the NO2B
level as well.

Here we encounter a subtlety.  If we were treating only a single nucleus (and
were not truncating operators), the reference state would be the ground state of
the RG-improved Hamiltonian $\tilde{H} \equiv H(s=\infty)$, with the exact
ground state energy as its eigenvalue. But our NME calculation requires the
ground states of both the initial and final nuclei, and so we adapt ideas from
Ref.~\cite{Stroberg:2017} by defining a reference \emph{ensemble} via the
density operator $\op{\rho} = c_I \ket{\Phi_I}\!\bra{\Phi_I} + c_F
\ket{\Phi_F}\!\bra{\Phi_F}$, where $\ket{\Phi_{I,F}}$ are the symmetry-restored
states with $J^{\pi} = 0^{+}$ obtained by projecting the lowest-energy
quasiparticle vacua for each nucleus onto good angular momentum and particle
number, and $c_I+c_F=1$.  Using the techniques of Refs.\
\cite{Kutzelnigg:1997,Mukherjee:1997}, we normal-order our Hamiltonian with
respect to this reference ensemble, again discarding explicit three-body pieces
to make the subsequent IMSRG evolution feasible.  Then, though neither
$\ket{\Phi_{I}}$ nor $\ket{\Phi_{F}}$ are eigenvectors of $\tilde{H}$ at
the end of the evolution, both are reasonable approximations to eigenvectors.
We tolerate loss of exact eigenstates at this stage of the calculation in order
to use a single set of transition operators, the lack of which complicated the
calculations in Ref.~\cite{Yao:2018wq}. 

So much for our procedure's first step.  The second improves the approximate
eigenstates of $\tilde{H}$, which at the conclusion of the flow are the original
reference states.  Because the IMSRG evolution incorporates \emph{dynamic}
correlations, involving only a few nucleons, into the RG-improved Hamiltonian
$\tilde{H}$, we should be able to obtain close-to-exact eigenvectors by admixing
into the states $\ket{\Phi_{I,F}}$ other states that differ only in their
collective parameters, to allow fluctuations in the deformation and pairing
condensate.  We thus use $\tilde{H}$ to perform a second set of projected-HFB
calculations that generate multiple (nonorthogonal) number-projected
quasiparticle vacua $\ket{\Phi_{ZN} (\vec{Q})}$ where $\vec{Q}=\{q_{\mu},
\phi\}$ encompasses the collective coordinates most important for spectra and
the NME \cite{Menendez:2016PRC}: quadrupole moments $q_{\mu}=\braket{\Phi_{ZN}
(\vec{Q})| \op{r}^2 \op{Y}_{2\mu} |\Phi_{ZN} (\vec{Q})}$ and an isoscalar
(proton-neutron) pairing amplitude $\phi=\braket{\Phi_{ZN} (\vec{Q})|
\op{P}^\dagger_0 + P_0 |\Phi_{ZN} (\vec{Q})}$.  Here $P_0^\dag$, defined
precisely in Ref.\ \cite{Hinohara:2014}, creates a correlated isoscalar pair.
We construct low-lying eigenstates by further projecting the $\ket{\Phi_{ZN}
(\vec{Q})}$ onto states with well-defined angular momentum,
$\ket{JMZN(\vec{Q}_i)}$, and superposing them using the GCM ansatz
\begin{equation}
\label{eq:GCM}
  \ket{\Psi^{JMZN}} = \sum_{\vec{Q}_i} F^{JZN}(\vec{Q}_i)
  \ket{JMZN(\mathbf{Q}_i)} \,.
\end{equation}
The weights $F^{JZN}(\vec{Q}_i)$ are determined by
minimizing the expectation value of the evolved Hamiltonian $\tilde{H}$, a
procedure that leads to the Hill-Wheeler-Griffin equation~\cite{Ring:1980}.
Since our approach involves a Hamiltonian, we
do not suffer from the spurious divergences and discontinuities that affect GCM
applications in nuclear DFT \cite{Bender:2009,Duguet:2009}.

\begin{figure}[t]
  \centering
  \includegraphics[width=4.2cm]{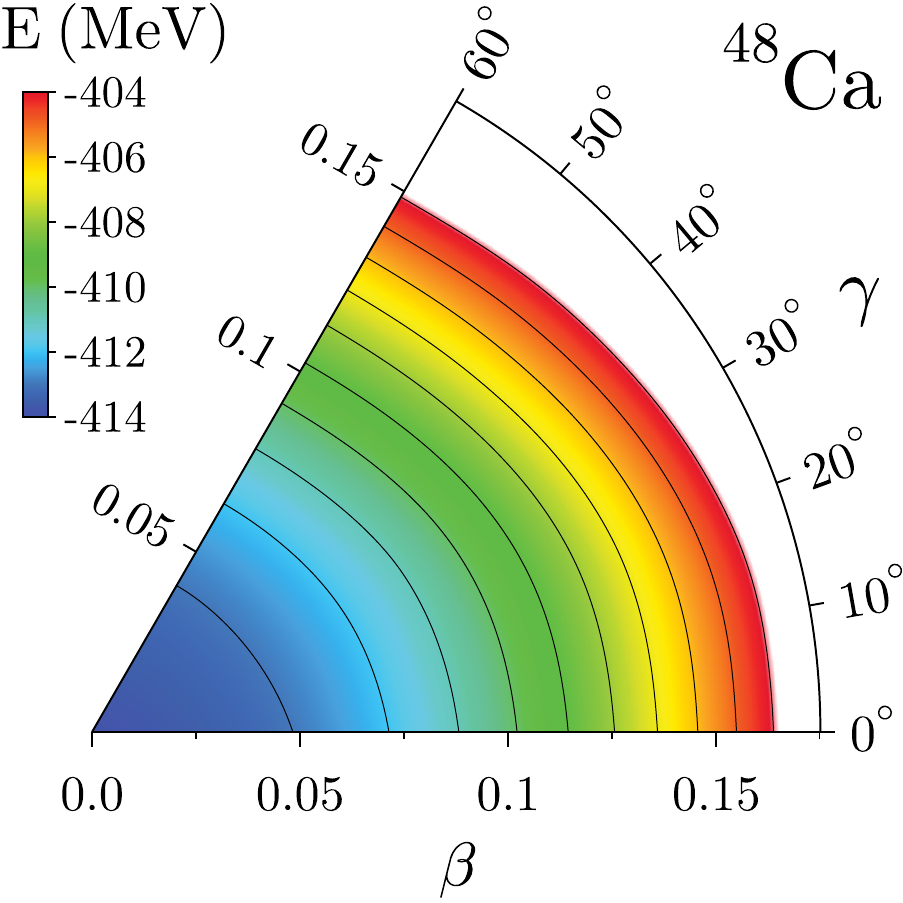} 
  \includegraphics[width=4.2cm]{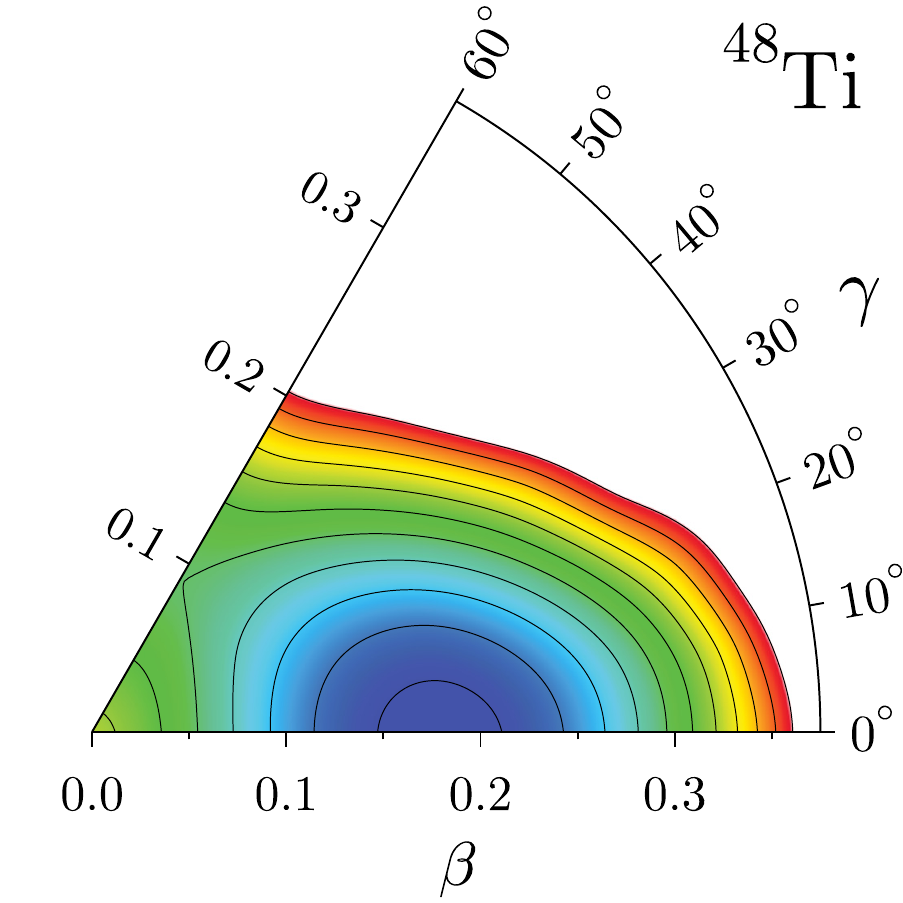}
  \caption{%
  The particle-number projected potential energy surfaces of $^{48}$Ca and
  $^{48}$Ti in the deformation $(\beta, \gamma)$ plane for the interaction
  EM1.8/2.0 with $\eMax=8$, $\hw=\SI{16}{\MeV}$ (see text). Neighboring contour
  lines are separated by \SI{1}{\MeV}.  } \label{fig:PES}
\end{figure} 

\paragraph{Results and discussion.}

Figure \ref{fig:PES} displays the ``potential energy surfaces,'' i.e.,
the expectation values $\braket{\Phi_{ZN} (\vec{Q}_{i})|\tilde{H}|\Phi_{ZN}
(\vec{Q}_{i})}$, for $^{48}$Ca and $^{48}$Ti.  The expectation value at each
deformation $(\beta, \gamma)$, where $\beta\equiv 4\pi/(3AR^2_0)
\sqrt{q_{0}^2+2q_{2}^2}$ with $R_0=1.2\,A^{1/3}\,\mathrm{fm}$ and
$\gamma\equiv\arctan{\sqrt{2}q_{2}/q_{0}}$, is an indication of the importance
of the corresponding state in our GCM wave functions.  The IMSRG-evolved
Hamiltonian $\tilde{H}$ used to construct the surface comes from the EM1.8/2.0
interaction, with $e_{\rm max} = 8$ and $\hbar\omega = 16$ MeV.  For convenience, 
we use the bare rather than the evolved quadrupole operators to define $\beta$ 
and $\gamma$; this convention has no effect on computed observables.  The figure 
shows that the energy of $^{48}$Ca is minimized for a spherical shape ($\beta=0,\gamma=0$), 
and that the energy of $^{48}$Ti has a similarly pronounced minimum at a prolate shape with 
$\beta\sim0.2$ and $\gamma=0$. The effect of triaxiality on the low-lying states of both nuclei
and on the NME is negligible.

We compute all observable quantities with the chiral interactions discussed
above, for a range of $\eMax$ and $\hw$ values (see Supplemental Material
for details.)  With EM1.8/2.0, which produces satisfactory
ground-state and separation energies through mass $A\sim80$
\cite{Simonis:2017,Holt:2019gm,Hagen:2016xe,Drischler:2019qf}, we obtain extrapolated
ground-state energies of -418.26 MeV and -422.27 MeV for $^{48}$Ca and
$^{48}$Ti, respectively.  Our calculation yields the correct ground-state
ordering, but our $Q_{\beta\beta}$=\SI{5.57}{\MeV} is somewhat larger than the
experimental $Q$ value, \SI{4.26}{\MeV}.  

\begin{figure}[t]
  \centering  
  \includegraphics[width=8cm]{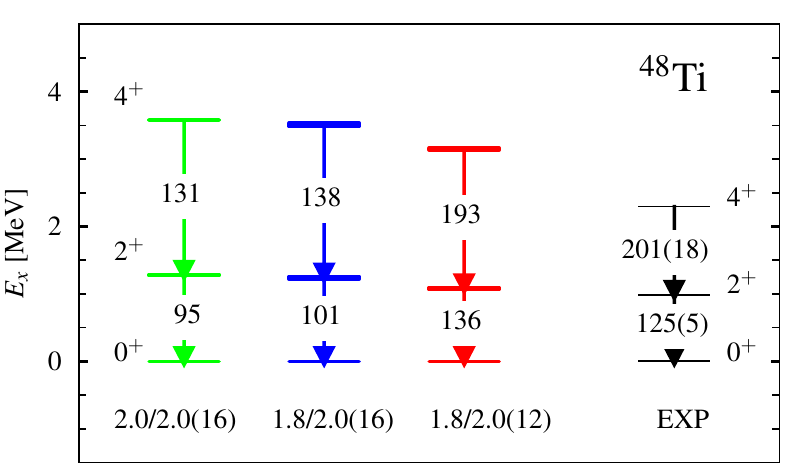} 
  \caption{The low-lying energy spectrum in $^{48}$Ti from the IMSRG+GCM
  calculation, with interactions and oscillator frequencies labeled
  EM$\lambda/\Lambda(\hbar\omega)$.  The rightmost column contains experimental data \cite{NNDC}. }
\label{fig:spectrum}
\end{figure}

Figure \ref{fig:spectrum} shows the low-lying states of $^{48}$Ti for the same
interactions. The spectrum is clearly rotational but slightly stretched, a
result of our focus on the ground state.  Importantly, however, we reproduce the
collective $B(\text{E2: }2_1^+ \to 0_1^+)$ reasonably well in all cases.  Other
\emph{ab initio} calculations severely underpredict $B(\text{E2})$'s 
\cite{Parzuchowski:2017ta,Stroberg:2019th},
which are more sensitive probes of wave functions than are energies; our success
is due to the explicit treatment of collectivity. The 
inclusion of noncollective configurations from isoscalar pairing, not shown in 
the figure, slightly compresses the spectra and changes the $B(\text{E2: }2_1^+ \to 0_1^+)$
by ~5-6\%, e.g., from \SI{101}{\elementarycharge\squared\fm\tothe4} to
\SI{96}{\elementarycharge\squared\fm\tothe4} for the EM1.8/2.0 interaction.  

\begin{figure}[t]
  \centering 
  \includegraphics[]{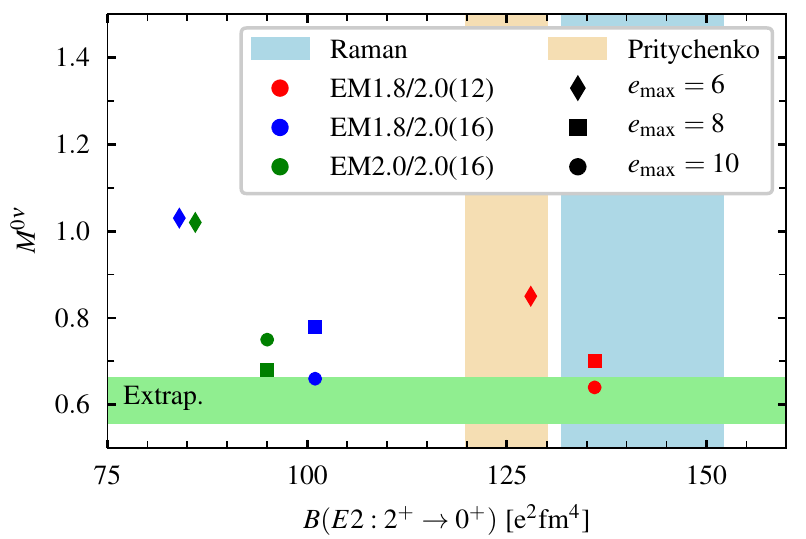} 
  \caption{The NME $M^{0\nu}$ versus the $B(\text{E2: }2_1^+ \to 0_1^+)$ value
  in $^{48}$Ti from IMSRG+GCM calculation, with different interactions,
  oscillator frequencies, and cutoffs. The vertical shaded areas indicate the
  experimental $B(\text{E2: }2_1^+ \to 0_1^+)$ values for $^{48}$Ti from Refs.
  \cite{Raman:2001,Pritychenko:2016}. The horizontal area represents the results
  ($0.61^{+0.04}_{-0.05}$) of extrapolation.} \label{fig:NME-BE2}
\end{figure}

The energies of the low-lying states are converged to within a few percent with
respect to the basis size. For example, the $2^+$
excitation energies in $^{48}$Ti obtained with EM1.8/2.0 or EM2.0/2.0 with
$\hbar \omega=16$ MeV change by no more than \SI{3}{\percent} from
$e_\text{max}=6$ through $e_\text{max}=10$ (also see Supplemental Material).  Regarding 
the transitions, we note first that the correction to the $E2$ operator from the 
IMSRG flow alters the
$B(E2)$ values by less than 10\%, suggesting that our collective reference ensemble
accounts for quadrupole correlations that caused large corrections in other
work \cite{Parzuchowski:2017ta}. Thus, we do not expect them to change significantly 
as the number of shells is increased (Fig.\
\ref{fig:NME-BE2} supports our expectation).  Surprisingly, even a drastic
change of the coefficients ($c_I, c_F$) specifying the contributions of
$^{48}$Ca and $^{48}$Ti to the reference ensemble from $(0.5, 0.5)$ to $(0.1,
0.9)$ changes the ground-state energy by a mere $100$-$200 \text{ keV}$,
excited-state energies by 5\% or less, and the $B(E2)$ by only 1\%.  

\begin{figure}[t]
 \centering
  \includegraphics[width=8cm]{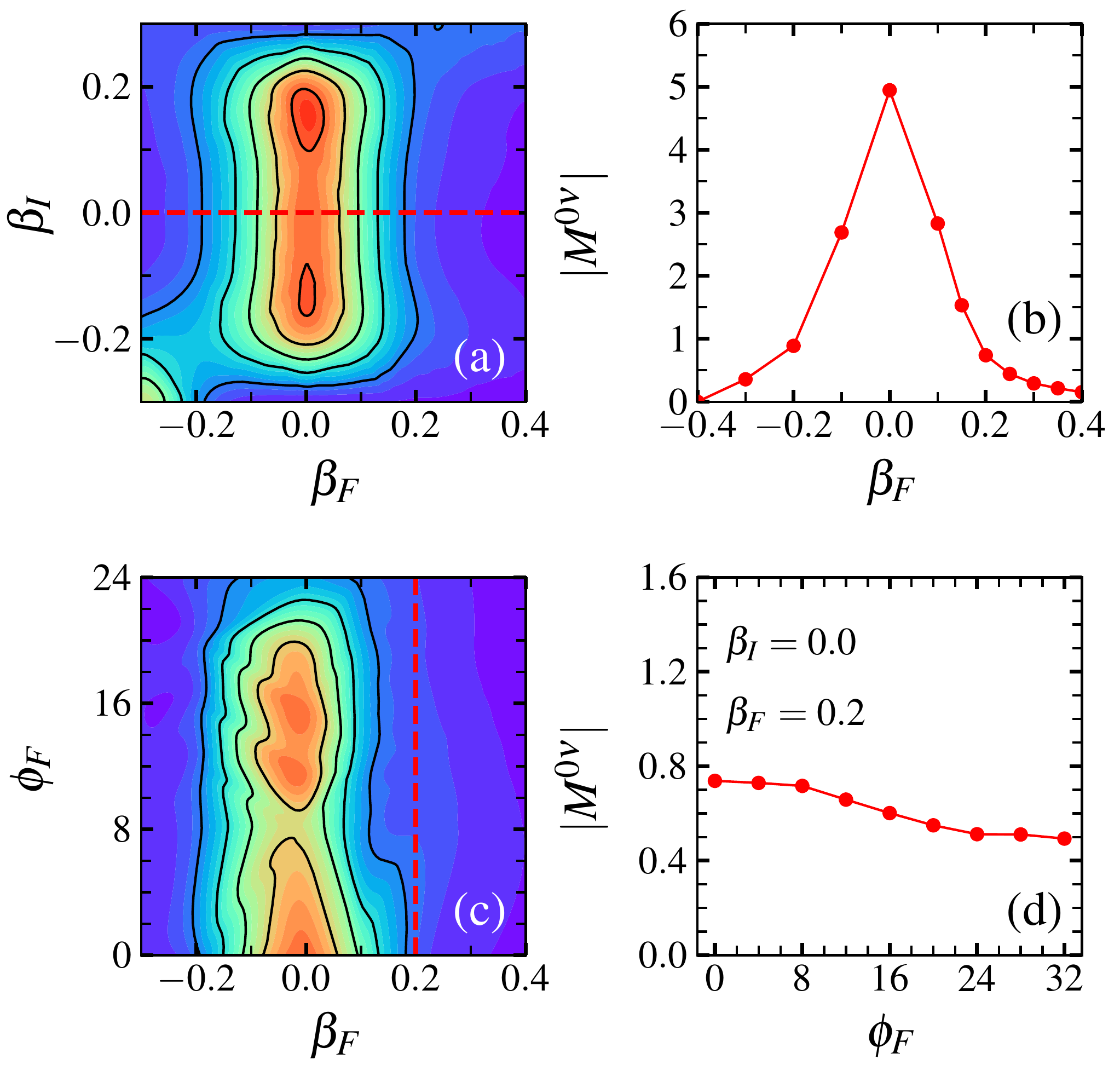} 
  \caption{ (a) Contributions to the NME $\abs{M^{0\nu}}$ in the $(\beta_I,
  \beta_F)$ plane (see text).  Neighboring contour lines here and in (c) are
  separated by $0.50$.  (b) The NME $\abs{M^{0\nu}}$ as a function of the
  quadrupole deformation parameter $\beta_{F}$ in $^{48}$Ti.  (c) Contributions
  to $\abs{M^{0\nu}}$ in the $(\beta_{F}, \phi_F)$ plane.  (d) The NME
  $\abs{M^{0\nu}}$ as a function of the proton-neutron pairing amplitude
  $\phi_{F}$ in $^{48}$Ti.} \label{fig:NME-all}
\end{figure}

\begin{table}[b] 
  \tabcolsep=6pt
  \caption{%
  The NME $M^{0\nu}$ for the decay
  $^{48}$Ca $\to$ $^{48}$Ti from the IMSRG+GCM calculation. The
  results labeled by */$\dag$ are from nonstandard reference ensembles
  with mixing weights $(1/3,2/3)$ and $(0.1, 0.9)$, respectively. For other cases the weights
 are $(1/2, 1/2)$.  }
  \begin{ruledtabular}
  \begin{tabular}{lcccc} 
  & & \multicolumn{3}{c}{NME}  \\ \cmidrule{3-5}
  Interaction & $\hbar\omega$ & $\eMax=6$ & $\eMax=8$ & $\eMax=10$ \\
  \midrule 
  EM1.8/2.0 & 12 & 0.85  & 0.70  & 0.64  \\
  EM1.8/2.0 & 16 & 1.03  & 0.78  & 0.66  \\
  \midrule
  EM2.0/2.0 & 16 & 1.02 & 0.68 & 0.75 \\
  \midrule
  EM1.8/2.0\rlap{$^{*}$}     & 16 &  & 0.81  & \\
  EM1.8/2.0\rlap{$^\dagger$} & 16 &  & 0.80  & \\
  \end{tabular}
  \end{ruledtabular}
  \label{tab:NME-axial} 
\end{table}

We turn next to the $0\nu\beta\beta$ NME, which we compute with the usual form
for the nuclear current operator \cite{Simkovic:2008}. 
We neglect newly discovered corrections to the decay operator in chiral EFT
\cite{Cirigliano:2018,Cirigliano:2018PRC} and many-body currents
\cite{Javier2011PRL,Wang2018}.  Figure \ref{fig:NME-all} displays NME
contributions from components with different 
values of the generator coordinates. These contributions are multiplied by the weight
functions $F$ in Eq.\ \eqref{eq:GCM} of both the initial and final states,
and then integrated to get the complete NME.  Figures \ref{fig:NME-all}(a) and 
\ref{fig:NME-all}(b) show that the
contributions hardly depend on the initial deformation $\beta_I$, as long as
that quantity is between $-0.2$ and $0.2$, but vary strongly with the final
deformation $\beta_F$.  
The significant average deformation of $^{48}$Ti means
that the NME will be suppressed. This result echoes the findings of DFT
calculations, which show strong quenching of NMEs between initial and final
states with different shapes \cite{Rodriguez:2010, Yao:2015}.
Figures \ref{fig:NME-all}(c) and \ref{fig:NME-all}(d) illustrate how the NME is
affected by isoscalar pairing, which was shown to be significant in
valence-space GCM calculations with empirical interactions
\cite{Menendez:2016PRC,Jiao:2017}.  The ground state of $^{48}$Ti is dominated
by configurations with $\beta_F \approx 0.2$, and at those values the isoscalar
pairing is significantly smaller than at $\beta_{F} = 0$.  Thus, its overall
effect on the NME is mild, as panel (d) demonstrates. (Isoscalar pairing in
$^{48}$Ca is negligible.)

For this reason and because GCM calculations with energy density functionals
find cancellations between isoscalar and isovector pairing fluctuations
\cite{Vaquero:2013}, we present results without the isoscalar-pairing amplitude
as a generator coordinate. Table \ref{tab:NME-axial} lists the full NMEs of both
nuclei with several interactions. The results in the largest model spaces are
relatively close to one another, ranging from $0.64$ to $0.75$.  All results are
summarized in Fig.\ \ref{fig:NME-BE2}. We find a noticeable but weak correlation
between the NME and the $B(\text{E2: } 2_1^+ \to 0_1^+)$ value in $^{48}$Ti,
especially in our largest model space. The weakness may reflect the relative
insensitivity of the NME to the correlations at large nucleon separations that
strongly affect $B(\text{E2})$'s. To extrapolate our results to even larger
model spaces, we perform a Bayesian fit of the parameters in the exponential
formula \cite{Basili:2019} $M^{0\nu}(e_{\rm Max})=M^{0\nu}(\infty) + a \exp(-b
 e_{\rm Max})$.  For the interaction EM1.8/20 with $\hbar \omega
=\SI{16}{\MeV}$, we obtain an extrapolated NME of $M^{0\nu}(\infty)=0.57$, with
an extrapolation uncertainty range of $(+0.08,-0.1)$; with
$\hbar\omega=\SI{12}{\MeV}$ we get $0.66$ with a range of $(+0.03, -0.10)$.
With the constraint that the extrapolated values for both oscillator frequencies
be equal, we obtain $M^{0\nu}(\infty)=0.61$ with an extrapolation uncertainty
range of $(+0.04, -0.05)$. For EM2.0/2.0, the oscillation with $e_{\rm max}$
prevents a similar analysis.  Our results carry additional uncertainties that 
will be investigated further by improving the IMSRG truncation and including
isoscalar and isovector proton-neutron pairing generator coordinates, 
as discussed above.  

All central values for our NMEs are 
near or below the predictions of phenomenological models \cite{Javier2009NPA,Simkovic:2013rw,Yao:2015,Iwata:2016}.
Only a very recent shell-model study \cite{Coraggio:2020} with 
perturbatively derived effective interactions and operators
yields an even smaller NME of 0.3. The small size of our NMEs seems 
to result from the interplay of valence-space and beyond valence-space correlations 
that our no-core approach is able to  capture.

Finally, we note a significant renormalization of the NME by the IMSRG flow.  
With the unevolved $0\nu\beta\beta$ operator, the NME for
EM1.8/2.0 at $\hbar\omega=16$ MeV and $e_{\rm max} = 8$ is 0.31 instead of 0.78.
A more detailed analysis shows that the flow incorporates the effects of pairing
in high-energy orbitals, greatly enhancing the $J=0$ contribution to the NME.
Breaking down the NME by the distance between the decaying nucleons, we confirm
that the largest contribution comes from a peak around 1 fm
(cf.~Ref.~\cite{Menendez:2009}), but contributions at larger distances are not
negligible (see Supplemental Material).

\paragraph{Conclusions.}

\emph{Ab initio} methods are essential for the calculation of $0\nu \beta
\beta$ matrix elements with real error estimates.   We have reported the first
\emph{ab initio} calculation of the spectrum and transition probabilities in a
deformed medium-mass nucleus, and significant progress in the \emph{ab initio}
computation of the NME for the $0 \nu \beta \beta$ decay of $^{48}$Ca. Our
approach, based on a novel method for deformed nuclei that combines the IMSRG
and GCM, has been validated in light nuclei against the no-core shell model 
(see Ref. \cite{Basili:2019} and a forthcoming paper) and allows a systematic 
exploration of theoretical uncertainties.  Here, we reproduce
the low-lying (collective) spectrum and $E2$ transitions of $^{48}$Ti satisfactorily.  
Our $B(E2)$ values are much improved compared to previous \emph{ab initio} calculations
\cite{Parzuchowski:2017ta},
but retain a significant dependence on $\hbar \omega$ even as we enlarge the model 
space.
This result supports the ideas behind our method: Collective correlations
that are omitted due to the IMSRG truncation can be captured by sophisticated reference 
states instead. Fortunately, the correlation between the NME and the $B(\text{E2})$ 
value is weak. 

Our best estimate of the NME with EM1.8/2.0 is $M^{0\nu}=0.61$; for EM2.0/2.0 it
would be a few percent larger. These values are near or below the
predictions of most phenomenological approaches \cite{Engel:2017}. Isoscalar
pairing can further reduce the NME by about \SI{17}{\percent}, but the effect might
be less when the calculation also includes isovector pairing fluctuations
\cite{Vaquero:2013}. 

To assign an overall error bar, we will implement improved IMSRG truncations
\cite{Hergert:2018th,Morris:2016xp,Hergert:2016jk}, explore additional collective correlations with the GCM, 
include the (small) effects of ``short-range correlations'' by evolving the 
$0\nu\beta \beta$ operator to the same resolution scale as the interaction, and 
consistently treat its contributions from higher orders in chiral EFT, including 
many-body currents \cite{Wang2018,Gysbers:2019df}.  We must also account for a 
leading-order contact operator \cite{Cirigliano:2018} whose coefficient is currently 
unknown.  We do not expect any of these steps (except perhaps the inclusion of the 
contact operator) to dramatically change the matrix element, and therefore plan to 
apply our new framework to heavier double beta decay candidate nuclei, such as 
$^{76}$Ge and $^{136}$Xe. 

\begin{acknowledgments}
\paragraph{Acknowledgements.} 

We thank A.~Belley, G.~Hagen, J.~D.~Holt, J.~Men\'endez, S.~Novario, T.~
Papenbrock, and S.~R. Stroberg for useful discussions and comments, and K.~Hebeler
for providing us with momentum space inputs and benchmarks during the
construction of our three-nucleon matrix elements. We also thank the Institute
for Nuclear Theory at the University of Washington for its hospitality during
the completion of this work, which is supported in part by the U.S.  Department
of Energy, Office of Science, Office of Nuclear Physics under Awards 
No.~DE-SC0017887, No.~DE-FG02-97ER41019, No.~DE-SC0015376 (the DBD Topical Theory
Collaboration) and No.~DE-SC0018083 (NUCLEI SciDAC-4 Collaboration). T.~R.~R. is 
supported in part by the Spanish MICINN under Contract No.~
PGC2018-094583-B-I00. Computing resources were provided by the Institute for
Cyber-Enabled Research at Michigan State University, the Research Computing
group at the University of North Carolina, and the U.S.~National Energy
Research Scientific Computing Center (NERSC), a DOE Office of Science User
Facility supported by the Office of Science of the U.S.~Department of Energy
under Contract No. DE-AC02-05CH11231.  
\end{acknowledgments}


%



 \clearpage

\section{Supplemental Material for: \emph{Ab Initio} Treatment of Collective Correlations and the Neutrinoless Double Beta Decay of \texorpdfstring{$^{48}$Ca}{48Ca}}

\author{J. M. Yao}
\email{yaoj@frib.msu.edu}
\affiliation{Facility for Rare Isotope Beams, Michigan State University,
East Lansing, MI 48824-1321}

\author{B. Bally}
\email{bbally@email.unc.edu}
\author{J. Engel}
\email{engelj@physics.unc.edu}
\affiliation{Department of Physics and Astronomy, University of North Carolina, Chapel Hill, North Carolina 27516-3255, USA}

\author{R. Wirth}
\email{wirth@frib.msu.edu}
\affiliation{Facility for Rare Isotope Beams, Michigan State University,
East Lansing, MI 48824-1321}

\author{T. R. Rodr\'{\i}guez}
\email{tomas.rodriguez@uam.es}
\affiliation{Departamento de Física Teórica y Centro de Investigación Avanzada en Física Fundamental, Universidad Autónoma de Madrid, E-28049 Madrid, Spain}

\author{H. Hergert}
\email{hergert@frib.msu.edu}
\affiliation{Facility for Rare Isotope Beams, Michigan State University,
East Lansing, MI 48824-1321}
\affiliation{Department of Physics \& Astronomy, Michigan State University,
East Lansing, MI 48824-1321}

\date{\today}



\maketitle



\section{Interactions}
In our calculations, we use interactions constructed in Ref.~\cite{Hebeler:2011}, 
which yield an empirically reasonable description of finite nuclei and infinite 
nuclear matter \cite{Simonis:2017,Holt:2019gm,Drischler:2019qf}.

The two-nucleon interaction is the next-to-next-to-next-to leading
order (N$^3$LO) potential with a momentum cutoff of $\Lambda=500$MeV/$c$ 
constructed by Entem and Machleidt \citet{Entem:2003}, abbreviated as 
EM in the present manuscript and elsewhere \cite{Hebeler:2011,Simonis:2017,Holt:2019gm,Drischler:2019qf}.  
The resolution scale of this interaction is lowered by means of a free-space
similarity renormalization group (SRG) evolution to $\lambda=\SI{1.8}{\per\fm}$ 
or $\lambda=\SI{2.0}{\per\fm}$, respectively \cite{Bogner:2010}. Following 
the strategy of Refs.~\cite{Nogga:2004il,Hebeler:2011}, the EM interaction 
is supplemented with a next-to-next-to leading (N$^2$LO order chiral 
three-nucleon force with cutoff $\Lambda=\SI{2.0}{\per\fm}\approx400$ MeV/$c$ 
whose low-energy constants $c_D$ and $c_E$ are fit to reproduce the triton 
binding energy and matter radius of $\nuc{He}{4}$ (the values of the LECs 
$c_1,3,4$ are the same as those used in the EM interaction). In this way, 
the authors of Refs.~\cite{Nogga:2004il,Hebeler:2011} aim to account for 
the evolution of an initial chiral 3NF to lower resolution, as well as any 
induced 3NFs from the evolution of the two-nucleon interaction.

\section{Structure properties}
  \begin{table}[b]
    \caption{%
        The structural properties of \nuclide[48]{Ti} from the IMSRG+GCM calculation with the chiral interactions, in comparison with available data.  All energies are in units of \si{\MeV} and $B(\text{E2})$ in \si{\elementarycharge\squared\fm\tothe4}. The results with $*/\dagger$ are from the calculations starting from the ensemble reference state with mixing weight $(1/3,2/3)/(0.1, 0.9)$. For other cases, $(1/2, 1/2)$ is used.
    }
    \begin{ruledtabular}
    \begin{tabular}{llllcc} 
        EM$\lambda/\Lambda(\eMax/\hbar\omega)$ & $E(0^+_1)$ & $E_x(2^+_1)$ &   $E_x(4^+_1)$ & $B(E2:2_1^+ \to 0_1^+)$   \\
        \midrule 
           EM1.8/2.0(6/16) &  -401.97   & 1.25 & 3.35   &  84 \\
           EM1.8/2.0(8/16) &  -418.22   & 1.24 & 3.57   & 101 \\
           EM1.8/2.0(10/16) &  -421.48  & 1.24 & 3.48    & 101  \\
           \midrule
           EM1.8/2.0(8/16)*        &  -417.96    & 1.30 & 3.62       & 99 \\
           EM1.8/2.0(8/16)$\dagger$ &  -418.32     & 1.32   & 3.71    &  100 \\
           \midrule
           EM1.8/2.0(6/12) &  -387.68   & 1.03 & 2.90   & 128 \\
           EM1.8/2.0(8/12) &  -409.65   & 1.06 & 3.10   & 136 \\
           EM1.8/2.0(10/12) &  -416.95   & 1.08 & 3.14  & 137 \\
           \midrule
            EM2.0/2.0(6/16) & -361.59 & 1.32 & 3.48 & 86 \\
           \color{black}
            EM2.0/2.0(8/16) &  -387.18 & 1.33 & 3.63 & 95  \\
            EM2.0/2.0(10/16) & -395.34 & 1.28 & 3.58 & 95  \\
          \midrule
          Exp. & -418.70 & 0.98 & 2.30 &  144 \cite{Raman:2001} \\
            &  &  &   & 125 \cite{Pritychenko:2016}\\
    \end{tabular}
    \end{ruledtabular}
    \label{tab:spectra}
\end{table}

Table \ref{tab:spectra} lists the detailed information on the low-lying states of \nuclide[48]{Ti} from IMSRG+GCM calculations by mixing axially deformed configurations. One can see that the low-energy spectra (except for the ground-state energies) and E2 transition strengths are converged rather well at $\eMax=10$.   

\begin{figure}
    \centering  
     \includegraphics[width=7cm]{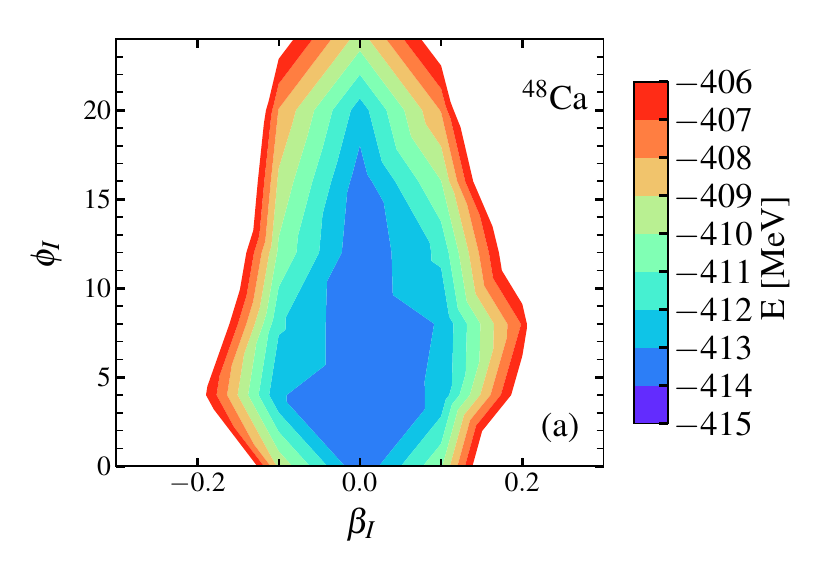} 
     \includegraphics[width=7cm]{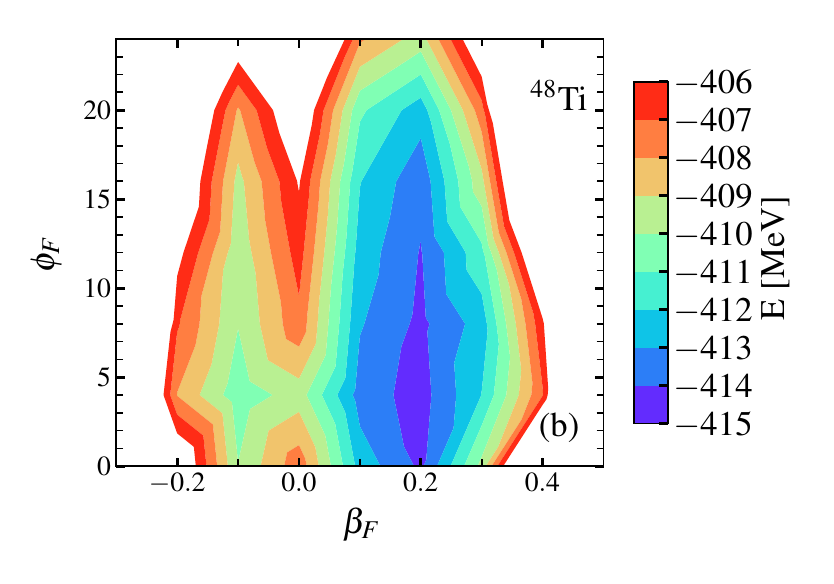} 
      \caption{The particle-number  projected  potential  energy  surfaces of \nuclide[48]{Ca} and  \nuclide[48]{Ti} in  the  $(\beta_2, \phi_{np})$ plane  at $e_{\rm max}=8$,  $\hw=16$ MeV.  The two neighbouring contour lines are separated by 1.0 MeV. }
        \label{fig:PES-bet2-phi}
\end{figure}

\begin{figure}
    \centering 
     \includegraphics[width=7cm]{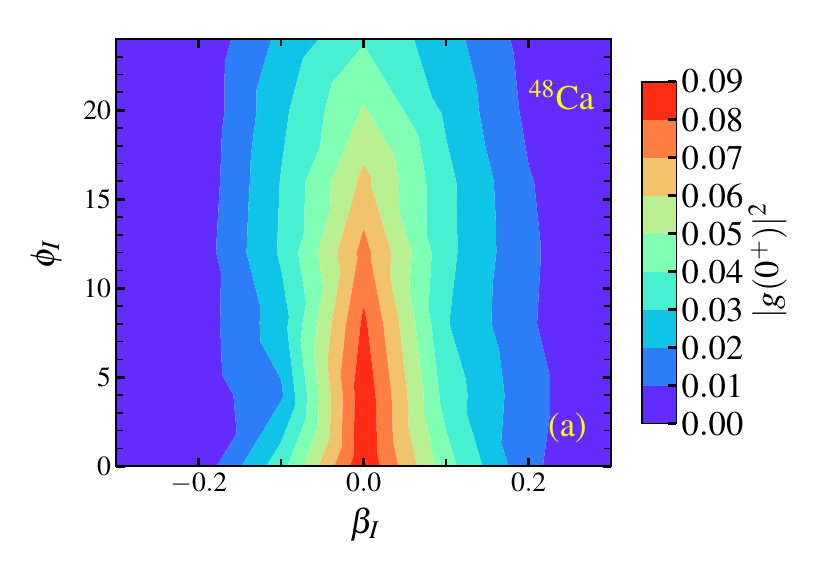} 
     \includegraphics[width=7cm]{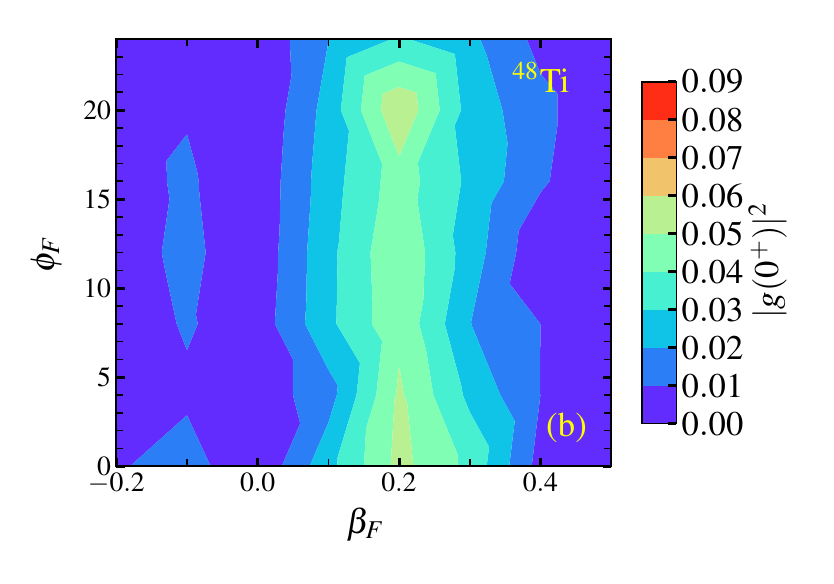} 
      \caption{The collective wave functions of the ground state for \nuclide[48]{Ca} and  \nuclide[48]{Ti} in  the  $(\beta_2, \phi_{np})$ plane  at $e_{\rm max}=8$,  $\hw=16$ MeV.  The two neighbouring contour lines are separated by 0.01. }
        \label{fig:wfJ0-bet2-phi}
\end{figure}

\begin{figure}
    \centering  
     \includegraphics[width=7cm]{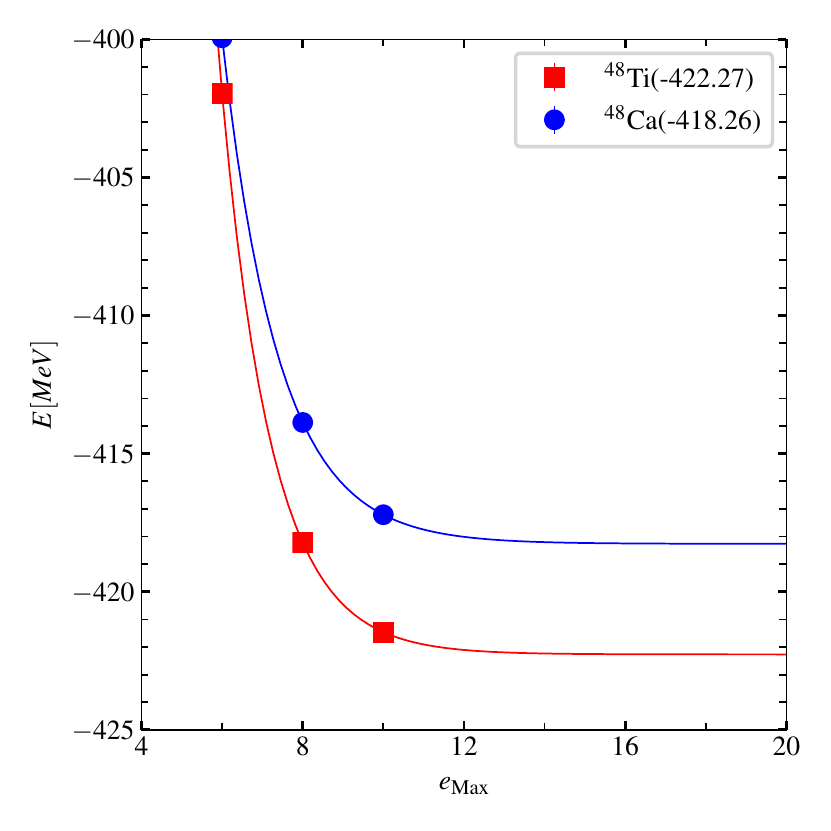}  
      \caption{The ground-state energy from the IMSRG+GCM calculation with the EM1.8/2.0 interaction with oscillator frequency $\hw=\SI{16}{\MeV}$ as a function of $\eMax$.  }
        \label{fig:energy-extrapolation}
\end{figure}

 Figure \ref{fig:PES-bet2-phi} shows the energy surfaces of \nuclide[48]{Ca} and  \nuclide[48]{Ti} in  the  $(\beta_2, \phi_{np})$ plane from the calculation using the EM1.8/2.0 interaction. It is seen clearly that the energy minimum is located at $\beta_2=0.0$ and $\beta_2=0.2$ for \nuclide[48]{Ca} and  \nuclide[48]{Ti}, respectively. Besides, the energy surface is rather soft along the neutron-proton isoscalar pairing amplitude $\phi_{np}$ in both nuclei around the energy minimum. Therefore, the wave functions of their ground states, which are relevant for the NME of the $0\nu\beta\beta$, are mainly concentrated along the the valley, as illustrated in Fig. \ref{fig:wfJ0-bet2-phi}. We note that the inclusion of $\phi_{np}$ degree-of-freedom only changes slightly the energies of low-lying states. For \nuclide[48]{Ca}, the ground-state energy is changed from -413.86 MeV to -413.87 MeV by the EM1.8/2.0 interaction with $\eMax=8$ and $\hw=16$ MeV. For \nuclide[48]{Ti}, it is changed from -418.22 MeV to -418.23 MeV. This effect decreases the excitation energy of $2^+$ state from 1.24 MeV to 1.17 MeV, closer to the data. Figure \ref{fig:energy-extrapolation} shows the convergence behavior of the ground-state energies as a function of the $\eMax$. The results are extrapolated with the exponential formula $E(e_{\rm Max})=E(\infty) + a \exp(-b \cdot e_{\rm Max})$. We find  $E(\infty)=-418.26$ MeV for \nuclide[48]{Ca} and $-422.27$ MeV for \nuclide[48]{Ti}.
 
\section{IMSRG-Evolved Neutrinoless Double Beta Decay Operator}

\begin{figure}
    \centering  
    \includegraphics[width=8cm]{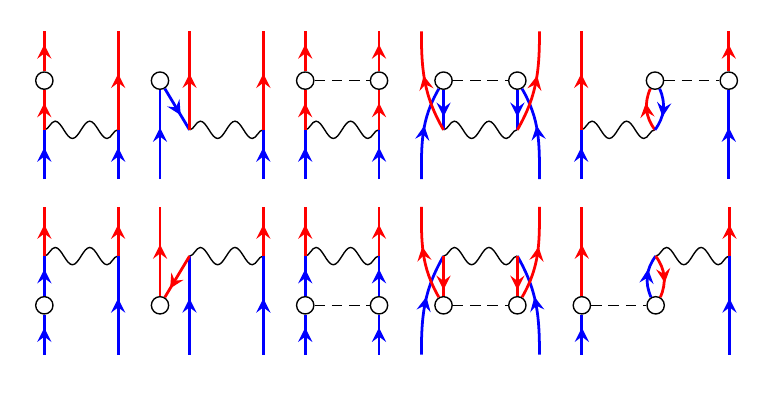}
       \caption{\label{fig:DBD}%
       Antisymmetrized Goldstone diagrams of the leading-order correction $\comm{\op{\Omega}}{ \op{O}^{0\nu}}$ from the IMSRG evolution to the $0\nu\beta\beta$ decay operator.
       Hollow dots mark insertions of one- or two-body parts of $\op{\Omega}$, wavy lines correspond to $\op{O}^{0\nu}$.
       The diagrams in the bottom row have to be subtracted.
       Diagrams containing two-body cumulants cancel and have been omitted.
       Formulas can be found in Ref.\ \cite{Yao:2018wq}.}
\end{figure}

In the IMSRG flow, the evolved $0\nu\beta\beta$ decay operator 
is calculated with the BCH formula,
\begin{equation} 
 \op{O}^{0\nu}(s) 
 = \op{O}^{0\nu} + [\op{\Omega}(s), \op{O}^{0\nu}]  + \dfrac{1}{2}[\op{\Omega}(s),[\op{\Omega}(s), \op{O}^{0\nu}] ] +\cdots
\end{equation}   
The leading-order correction $[\op{\Omega}(s), \op{O}^{0\nu}]$ is illustrated schematically with Goldstone diagrams in Fig.~\ref{fig:DBD}.  The correction from this term to the operator goes to all the terms in the BCH formula.  We find that the contribution from the one-body part (first two columns in Fig.~\ref{fig:DBD}) of $\Omega(s)$ to the commutator $[\op{\Omega}(s), \op{O}^{0\nu}]$ is negligible. The contribution of the diagrams in the third and four columns enhances the two-body matrix elements, while that of the last column quenches the matrix elements.

\section{Configuration-Dependence of the Nuclear Matrix Element}
 
  \begin{figure}
    \centering  
     \includegraphics[width=7cm]{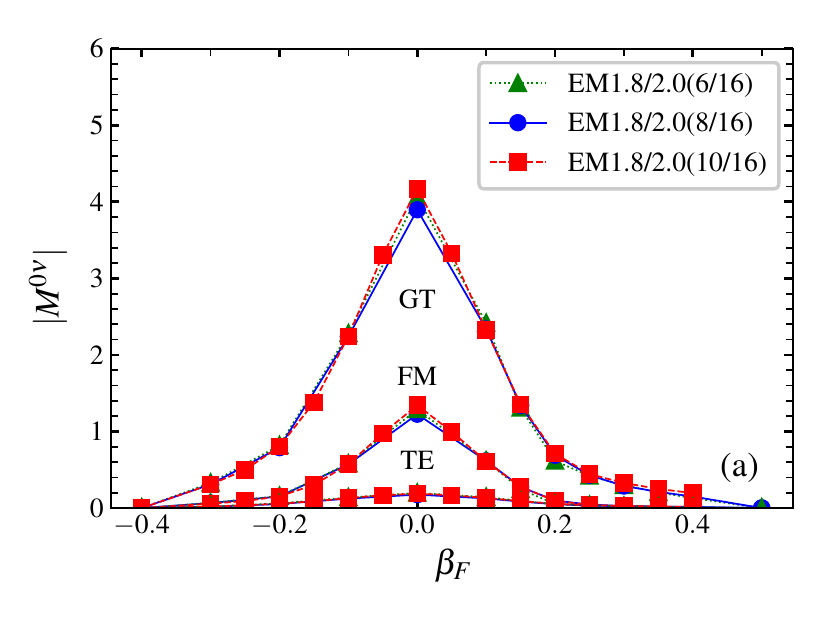} 
     \includegraphics[width=7cm]{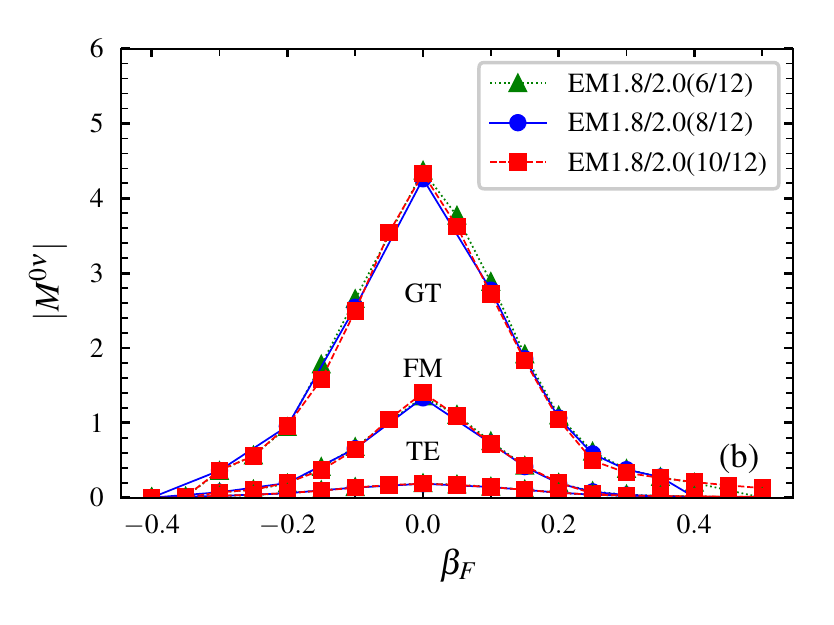} 
     \includegraphics[width=7cm]{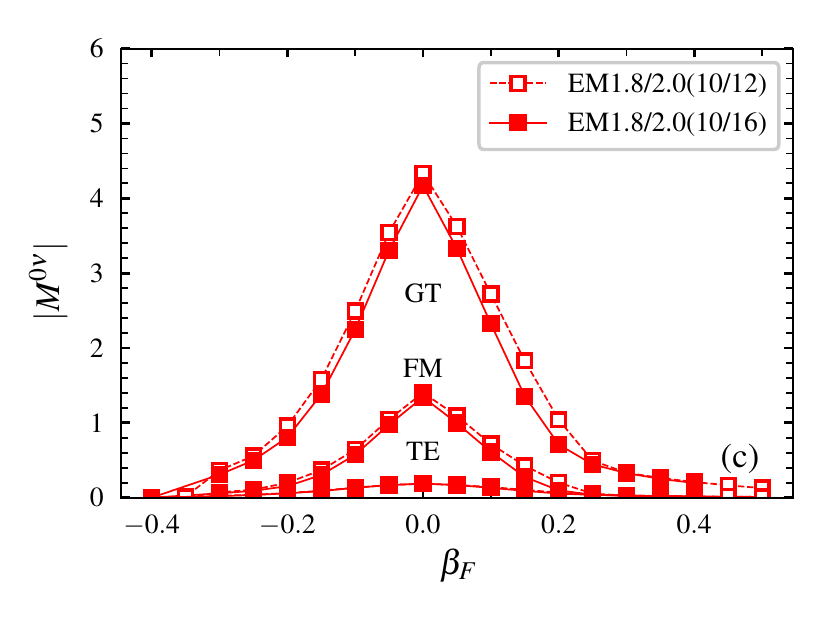} 
      \caption{ (a) The normalized NME $M^{0\nu}$ from the spherical state of \nuclide[48]{Ca} to  different axially deformed state of \nuclide[48]{Ti} from the IMSRG+GCM calculation using EM1.8/2.0($\eMax/16$) with $\eMax=6, 8,$ and 10, respectively; (b) Same as (a) but by the EM1.8/2.0($\eMax/12$) ; (c) Comparison of the results by EM1.8/2.0($10/12$) and EM1.8/2.0($10/16$). }
        \label{fig:NME-beta2}
\end{figure} 

Figure \ref{fig:NME-beta2} displays that the NME is very sensitive to the quadrupole deformation of \nuclide[48]{Ti}, but not much to the $\eMax$ for a given deformed state with the same $\beta^{(F)}_2$ value. Moreover, one can see that the NME by $\hbar\omega=12$ MeV is systematically larger than that by $\hbar\omega=16$ MeV. However, as we find, the $E2$ transition strengths by the former is also systematically larger than  the later. Because of negative correlation between the NME and $E2$ transition strength, these two interactions predicts a similar value for the NME in the final IMSRG+GCM calculation.

  \begin{figure}
    \centering 
     \includegraphics[width=5cm]{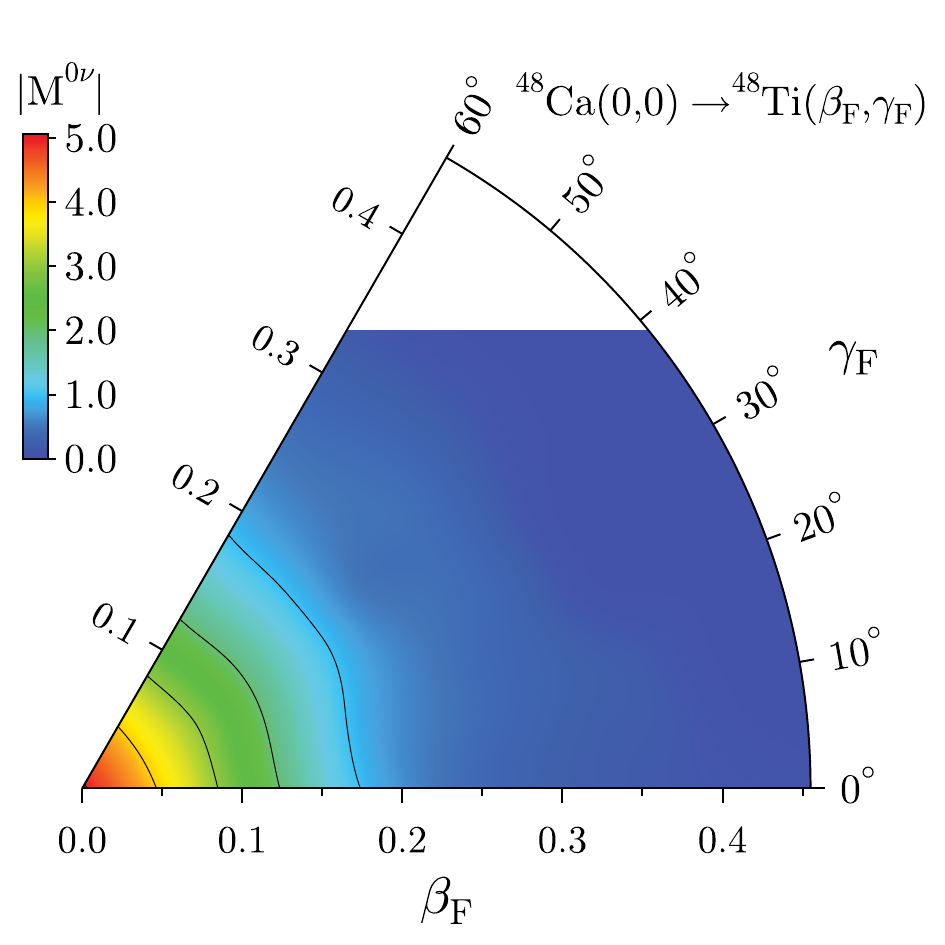}    
      \caption{The configuration-dependence of the NME  in the $(\beta_{2}, \gamma)$ plane  by the EM1.8/2.0(8/16) calculation. }
        \label{fig:NME-gammma}
\end{figure} 

  \begin{figure}
    \centering 
     \includegraphics[width=7cm]{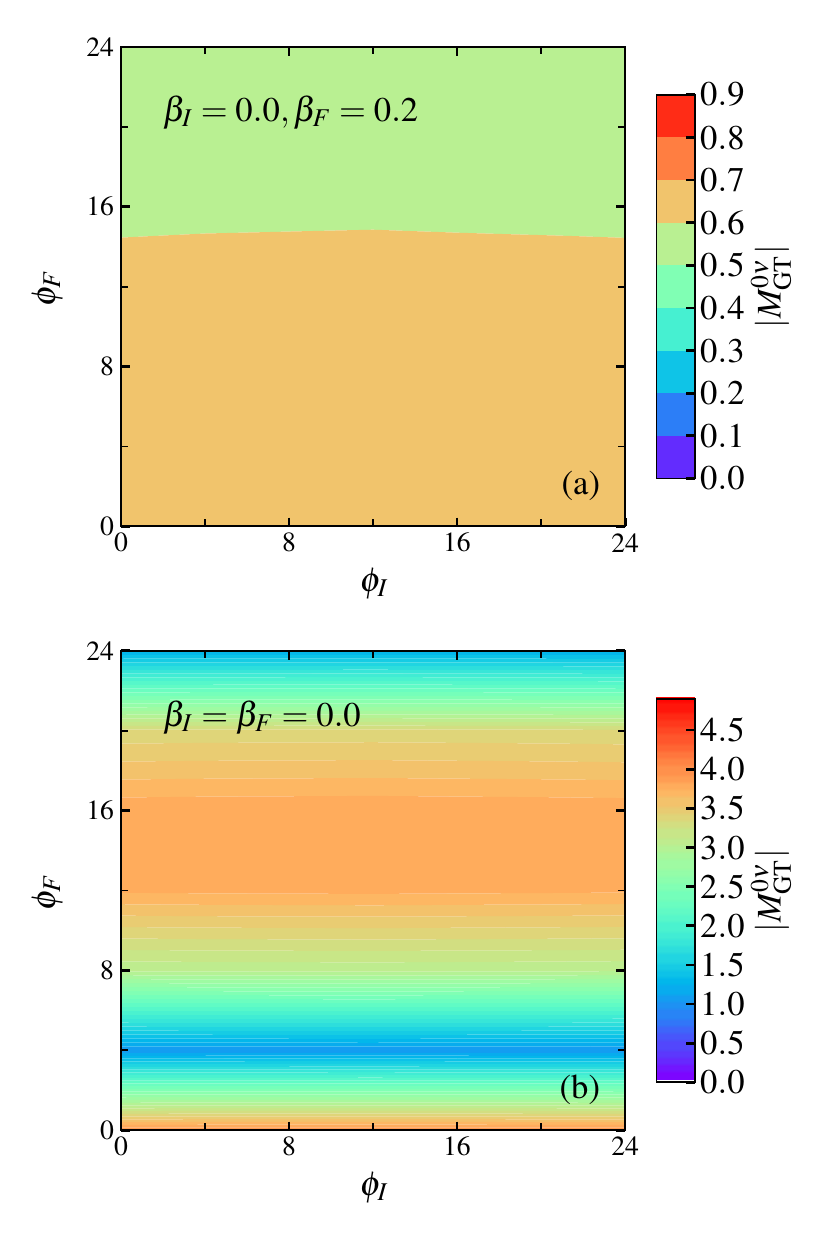}   
      \caption{The configuration-dependence of the NME  in the $(\phi^{(I)}_{\rm np}, \phi^{(F)}_{\rm np})$ plane  at $e_{\rm max}=8$,  $\hw=16$ MeV. }
        \label{fig:NME-phinp}
\end{figure} 

\begin{table}[b] 
    \caption{%
        The nuclear matrix element $M^{0\nu}$ for the $0\nu\beta\beta$ \nuclide[48]{Ca} $\to$ \nuclide[48]{Ti} from the IMSRG+GCM calculation without ($w/o$) and with($w/$) the neutron-proton isoscalar pairing fluctuation using the EM1.8/2.0 interaction and $\hw=16$ MeV. The quenching factor $q$ is defined as  the ratio of the difference in the NMEs to that without $np$ isoscalar pairing.  }
    \begin{ruledtabular}
    \begin{tabular}{cccc} 
      $\eMax$ &   $M^{0\nu}(w/o)$ & $M^{0\nu}(w/)$ & $q$ \\
        \midrule  
            6  &   1.03  &    0.86& 16.5\% \\
            8  &   0.78 &    0.65 & 16.7\%  \\
    \end{tabular}
    \end{ruledtabular}
    \label{tab:NME-axial-np} 
\end{table}

Figure \ref{fig:NME-gammma} shows that the NME is insensitive to the triaxial $\gamma$ deformation of \nuclide[48]{Ti}. The inclusion of triaxiality in the GCM calculation is thus expected to be negligible. Figure \ref{fig:NME-phinp} displays the sensitivity of the NME to the neutron-proton isoscalar pairing amplitude $\phi_{\rm np}$ in both initial and final nuclei for a given quadrupole deformation parameter.  It is seen that the NME is almost independent of  $\phi^{(I)}_{\rm np}$ in \nuclide[48]{Ca}, but may vary rapidly with $\phi^{(F)}_{\rm np}$ in \nuclide[48]{Ti}, depending on the $\beta^{(F)}_2$.

Table \ref{tab:NME-axial-np} lists the NME from the IMSRG+GCM calculation with neutron-proton isoscalar pairing fluctuation with $\eMax=6$ and 8. Compared with the value from the same calculation but without the isoscalar pairing fluctuation, this value is smaller by about 17\% in both cases.

\section{Distance- and Angular-Momentum Dependence of the NME}

We can analyze the NME further by breaking it down into contributions from different distances between the decaying nucleons and partial waves. For the former, we introduce the distribution $C^{0\nu}(r_{12})$, defined by 
\begin{equation}
    M^{0\nu} = \int^\infty_0 \dd r_{12}\, C^{0\nu}(r_{12}) \,,
\end{equation}
with $r_{12}=|\vec{r}_1 - \vec{r}_2|$ the relative distance between the two neutrons that are transformed into protons. 

\begin{figure}
    \centering 
     \includegraphics[width=7cm]{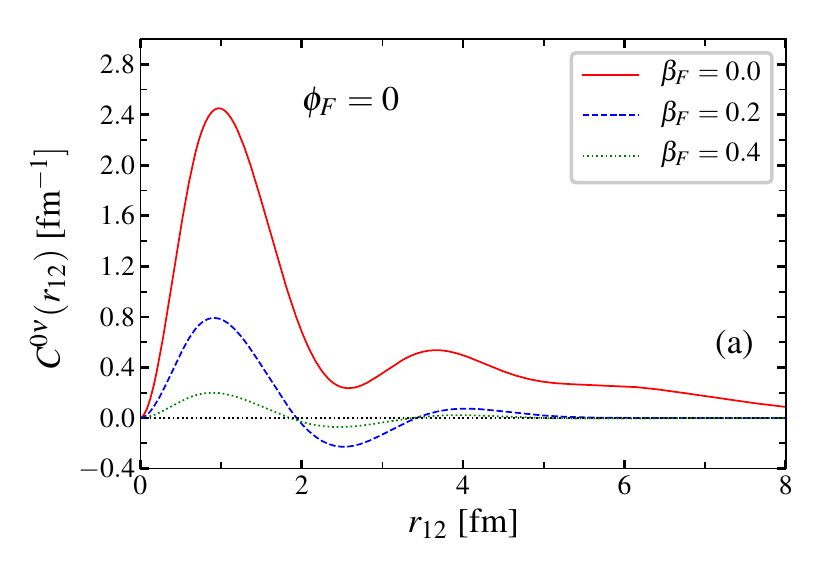} 
     \includegraphics[width=7cm]{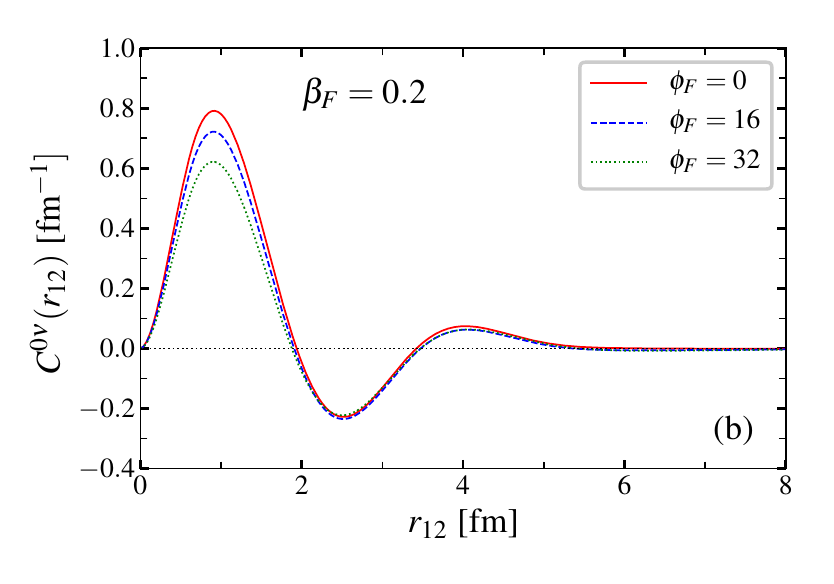} 
    \caption{The distribution of $C^{0\nu}(r_{12})$ as a function of the relative coordinate $r_{12}$ corresponding to the transition from spherical \nuclide[48]{Ca} to different configuration of \nuclide[48]{Ti} by EM1.8/2.0(8/16) distinguished with different values of quadrupole deformation parameter $\beta_2$ and neutron-proton isoscalar pairing amplitude $\phi_{np}$. }
     \label{fig:NME-r12-config}
\end{figure} 

Figure \ref{fig:NME-r12-config} shows distributions of the NME corresponding to the transition from spherical \nuclide[48]{Ca} to \nuclide[48]{Ti} with different quadrupole deformation and isoscalar pairing amplitude. One can see that the  quadrupole correlation quenches the NME in both the long-ranged and short-ranged region. In contrast, the isoscalar pairing quenches the NME mainly in the short-ranged region. The distributions with deformed final states are qualitatively similar to those obtained in phenomenological shell-model calculation of Ref.\ \cite{Menendez:2009}, featuring the appearance of a robust node at $r_{12}\approx\SI{2.0}{\fm}$.

\begin{figure}
    \centering 
     \includegraphics[width=7cm]{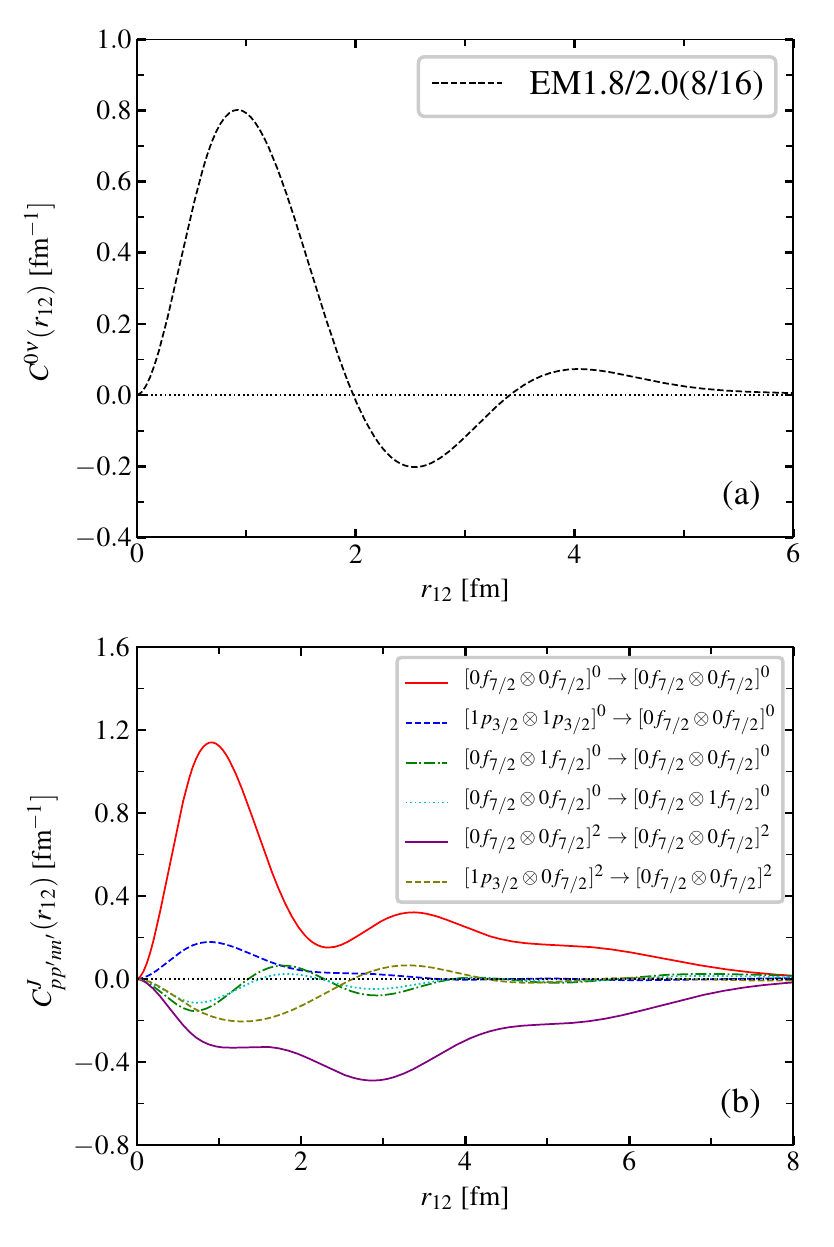}  
         \caption{ (a) The distribution of $C^{0\nu}(r_{12})$ as a function of the relative coordinate $r_{12}$ from the IMSRG+GCM calculation by EM1.8/2.0(8/16). (b) The contribution from the first six largest two-body transition densities.  }
     \label{fig:NME-r12-J}
\end{figure} 

 We decompose the $C^{0\nu}(r_{12})$ further into different $J$ components
\begin{align}
C^{0\nu}(r_{12}) &= \sum_{\substack{p\leq p' \\ n\leq n'}}\sum_J C^J_{pp'nn'}(r_{12}),
\shortintertext{where} 
 C^J_{pp'nn'}(r_{12})
  &= \dfrac{(2J+1)}{\sqrt{(1+\delta_{pp'})(1+\delta_{nn'})}} \notag\\
    &\phantom{{}={}} \times \braket{(pp')J| \bar{\op{O}}^{0\nu} (r_{12}) |(nn')J} 
 \rho^J_{ pp' nn'},
\end{align}
 with $\braket{(pp')J| \bar{\op{O}}^{0\nu} (r_{12}) |(nn')J}$ being the normalized IMSRG evolved two-body transition matrix element in harmonic-oscillator basis and the two-body transition density
 \begin{equation}\rho^J_{pp'nn'}
 = -\dfrac{1}{\sqrt{2J+1}}\braket{\Psi(\nuclide[48]{Ti})| [\op{a}^\dagger_{p}\op{a}^\dagger_{p'}]^J
 [\op{\tilde a}_{n}\op{\tilde a}_{n'}]^J| \Psi(\nuclide[48]{Ca})}.
 \end{equation}

  Figure~\ref{fig:NME-r12-J}  displays the $r$-dependence of the $C^J_{pp'nn'}(r_{12})$ from the first six largest two-body transition densities. It is seen that the cancellation between $J=0$ and $J=2$ components is the main mechanism responsible for the node around $r_{12}=\SI{2.0}{\fm}$, before and after which point the contribution to the total NME is opposite. It is the large negative contribution from the $J=2$ components arising from the strong quadrupole collectivity in \nuclide[48]{Ti} quenches the NME strongly.

 \section{Extrapolation to Infinite Model Space}

  \begin{figure}
    \centering 
     \includegraphics[width=7cm]{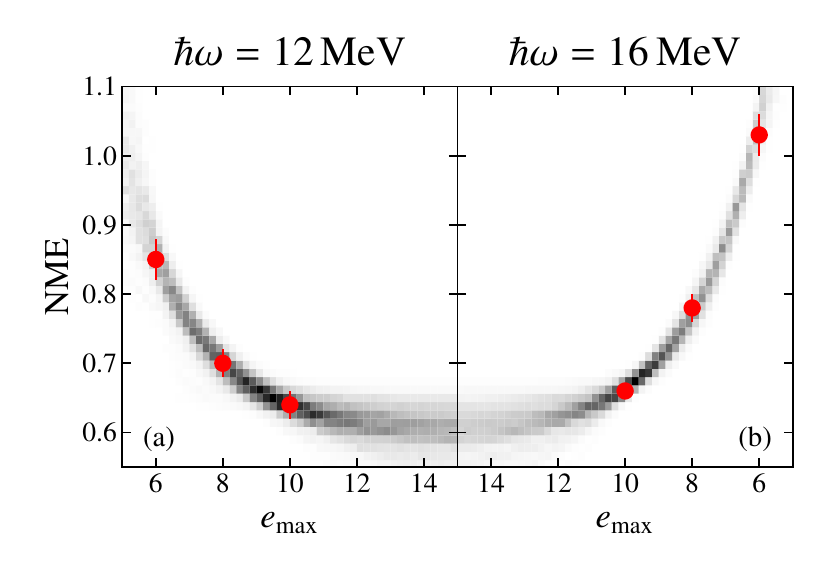}
       \caption{The NME $M^{0\nu}$ from the IMSRG+GCM calculation with the EM1.8/2.0 interaction with oscillator frequencies $\hw=\SI{12}{\MeV}$ and $\hw=\SI{16}{\MeV}$ as a function of $\eMax$.
       The histogram in the background shows the density of 500 realizations of the extrapolating curve drawn from its posterior distribution.
       Both curves are constrained to yield the same extrapolated value.}
        \label{fig:NME-fit}
\end{figure}

To get a reliable estimate of the NME extrapolated to infinite model space and its error in the presence of uncertainty in the calculation results, we perform Bayesian inference on the parameters of the exponential formula $M^{0\nu}(e_{\rm Max})=M^{0\nu}(\infty) + a \exp(-b \cdot e_{\rm Max})$. The results are displayed in Figure \ref{fig:NME-fit}. We use normal priors for the extrapolated value ($\mu = 0.5$, $\sigma=1$) and for the value at $\eMax=6$, where the expectation value is set to the central value of the calculation result and we set $\sigma=0.2$ to allow for some variation in the extrapolating curve.
The prior for the decay constant is a truncated normal distribution ($\mu=0.5$, $\sigma=2$) with the probability density for values less than zero removed in order to get a decaying exponential.
We assume Gaussian uncertainties on the resulting values of the NME.

First, we perform inference separately on the $\eMax$ sequences for both oscillator frequencies.
The uncertainty on the value of the NME at finite $\eMax$ allows for two solutions, one with a large decay constant and an extrapolated value in the region of $0.6$, and one with a small decay constant and very low extrapolated value.
This leads to a posterior distribution of the extrapolated value with a very long tail ranging into negative NME values.
However, the bulk of the probability is concentrated in a rather small interval.
For these reasons, we use the mode of the posterior as a point estimate and the \SI{68}{\percent} highest posterior density credible interval to define its uncertainty. For $\hw=\SI{16}{\MeV}$, we get an extrapolated value of $0.57$  with an extrapolation uncertainty range of $(+0.08,-0.1) $, the result for $\hw=\SI{12}{\MeV}$ is $0.66$ with extrapolation uncertainty range $(+0.03,-0.1)$.
To improve upon this and to get rid of the long tails in the posterior distribution, we also perform a combined fit where the extrapolated value is constrained to be the same for both oscillator frequencies while the other parameters are independent.
The inference on the combined data set yields a much more narrow distribution of the extrapolated value and reduces the probability of the slowly decaying solutions.
The extrapolated value in this case is $0.61$ with an extrapolation uncertainty range $(+0.04,-0.05)$.

\end{document}